\documentclass[dvips]{article}

\usepackage{icrc2011}

\title{Measurement of the cosmic electron plus positron spectrum with the MAGIC telescopes}

\newcommand{\etal}{\MakeLowercase{\textit{et al. }}} 
\shorttitle{author \etal paper short title}

\authors{D. Borla Tridon$^{1}$, P. Colin$^{1}$, L. Cossio$^{2}$,
  M. Doro$^{3}$, V. Scalzotto$^{4}$, \\ on behalf of the MAGIC Collaboration}
\afiliations{$^1$Max-Planck-Institut f\"ur Physik, Munich, Germany\\
  $^2$Universit\`a di Udine and INFN sez. di Trieste\\$^3$Universitat
  Aut\`onoma de Barcelona, Barcelona, Spain\\$^{4}$Universit\`a di Padova and INFN sez. di Padova}
\email{dborla@mppmu.mpg.de}

\abstract{Cosmic electrons with energies in the TeV range lose their
  energy rapidly through synchrotron radiation and inverse Compton
  processes, resulting in a relatively short lifetime
  ($\sim$\,10$^{5}$\,years). They are only visible from comparatively
  nearby sources ($<$\,1\,kpc).  Unexpected features in their spectrum
  at a few hundreds GeV, as measured by several experiments (ATIC,
  Fermi and H.E.S.S. among others), might be caused by local sources
  such as pulsars or by dark matter annihilation/decay. In order to
  investigate these possibilities, new measurements in the TeV energy
  region are needed. Since the completion of the stereo system, the
  MAGIC Cherenkov experiment is sensitive enough to measure the cosmic
  electron flux between a few hundred GeV and few TeV. The electron
  signal has to be extracted from the overwhelming background of
  hadronic cosmic rays estimated through Monte Carlo simulations. Here we present the first results of the cosmic electron spectrum measured with the MAGIC telescopes.}

\begin{document}
\maketitle

\section{Introduction}
TeV cosmic electrons and positrons (hereafter electrons) lose energy
mainly due to inverse Compton
processes and due to synchrotron radiation caused by the weak cosmic magnetic field. 
These losses limit the distance at which they can be
observed. Therefore, their energy spectrum can give us clues about
their origin, providing useful information on the nearby Universe.\\
Recently, several measurements of cosmic electrons have shown some features in their spectra, which have excited
astrophysicists in the field. Many interpretations in terms of dark
matter scenarios \cite{dm} or astrophysical sources such as pulsars \cite{pulsar} or supernova
remnants \cite{sn} are invoked.\\
Until recently, energy spectra measurements of electrons were obtained by balloon and satellite experiments. At TeV energies, however, the sensitivity of these instruments is insufficient due to their small sizes and short exposure time of flight.
Nowadays, ground-based Cherenkov telescopes, with their large
collection areas and good sensitivity, represent an excellent tool
for measuring high energy cosmic electrons, via the
indirect observation of the air showers that charged particles
generate in the atmosphere. In 2009 H.E.S.S. measured
the electron spectrum from 300\,GeV up to $\sim$\,4\,TeV \cite{hesshigh}.
MAGIC, with its two largest-dish Cherenkov telescopes
world-wide, is now one of the most suited experiments that can
contribute to this measurement. It has the potential to overlap with the
energy range of the other experiments (Fermi \cite{fermielectron2}, ATIC \cite{atic} and H.E.S.S. \cite{hesslow} \cite{hesshigh} among
the others), confirming previous measurements of the electron spectrum to TeV
energies.\\
The determination of the cosmic electron spectrum is pursued through a non-standard
analysis, which is instead optimized for the reconstruction of images
from $\gamma$-ray showers. Nonetheless, as in the case of $\gamma$-rays, electrons
events are largely sub-dominant, and overwhelmed by a much larger
background of hadronic events and also $\gamma$-ray events.
While $\gamma$-rays are not deflected by the magnetic fields and thus
their arrival direction points directly to their source, electrons are
isotropically diffused. Therefore, the identification of charged
particles cannot be done through the arrival direction information, but only via the shape of the image, which is produced by the air showers in the atmosphere and recorded by the telescopes. Data are modelled with simulations of electron showers and the background is rejected by applying selection criteria. 
Since this method does not separate electrons from gammas, a small
contamination from diffuse gammas is expected. Moreover, Cherenkov telescopes do not discriminate between the charge of the particles, thus the presented measurement includes the contribution of both electrons and positron.\\
\section{The MAGIC telescopes}
MAGIC (\textit{Major Atmospheric Gamma Imaging Cherenkov}) is a
stereoscopic system of two imaging atmospheric Cherenkov telescopes
(IACT). Among the operational IACTs, it has the world-largest dishes with 17m diameter each. They are located at the Observatory of the Roque de los Muchachos on the Canary Island of La Palma (28.75$^o$\,N, 17.86$^o$\,W, 2200\,m a.s.l.).\\
Since fall 2009 MAGIC is fully operational in stereoscopic mode.
The two telescopes are designed to detect very high energy (VHE)
$\gamma$-rays in the energy range from 50 GeV to tens of TeV and can
also detect electrons in the same energy range \cite{borla2}.
\section{Analysis}
The data used in this analysis come from observations of selected extragalactic sky
areas (in which no $\gamma$-ray emission from
astrophysical objects has been found) in order to minimize the contribution from the
diffuse $\gamma$-ray emission which comes mainly from the Galactic
plane. A priori, a search for a gamma-point source has been
carried out to demonstrate that no gammas contaminated the sample.
The data were extracted from observation carried out in December 2009, June 2010, October 2010 and
November 2010. Data passing quality selection criteria, with zenith
angles between 14$^o$ and 27$^o$, have been used in the analysis, providing
a total amount of $\sim$\,14\,hours. The images are cleaned using a
threshold of 6 (9) photoelectrons (phes) for
core pixels and 3 (4.5) phes for boundary pixels has been applied for
the MAGIC-I (MAGIC-II) telescope data respectively.
Higher cleaning for the MAGIC-II telescope data is needed because
of a higher photon to photoelectron efficiency and a higher noise level in the read-out chain
compared  to MAGIC-I read-out. Beside, the arrival
time in each pixel belonging to shower image can deviate at most of 4.5\,ns from the
shower core arrival time. Therefore we set a maximum time difference between
adjacent pixels to be less than 1.5\,ns.\\ \\ The most important issue in the electron
analysis is the electron/hadron separation and the rejection of the
hadronic background. A classification method (Random Forest - RF) \cite{rf} is used to compute the
\textit{Hadronness} parameter, which spans from 0 to 1 and gives the
degree of hadron-like. A classification tree of the RF is trained with
a sample of Monte Carlo (MC) electrons and a sample of MC protons. The input parameters of the RF are
the Hillas parameters \cite{hillas} of the shower images (\textit{Size}, \textit{Width},
\textit{Length} for both the telescopes), the \textit{Impact} parameter of the two telescopes and the
reconstructed \textit{Height} of the shower maximum. A cut on the
number of \textit{Islands} parameter has been applied for both the telescopes.
\\ \\ The reconstruction of the energy of each event is done via a
look-up-table obtained also from MC events, based on
\textit{Size}, \textit{Impact}, \textit{Height} and \textit{Zenith}
angle. The mean energy resolution is below 20\,$\%$ in the energy
range between 100\,GeV and 2\,TeV.
\\ \\ The signal of the diffuse cosmic electrons is determined from the \textit{Hadronness} parameter distribution (the result of the
RF). In the distribution of the \textit{Hadronness} the diffuse electron signal peaks around
\textit{Hadronness}\,=\,0 (electron-like), while the hadronic background around
\textit{Hadronness}\,=\,1 (hadron-like). 
In order to determine the electron spectrum, the following procedure is applied:
\begin{enumerate}
\item Apply selection cuts.
\item Determine the \textit{Hadronness} distribution for data and background, define a signal region and normalize the two distributions by the number of events in an optimized non signal region.
\item Count events in the signal region ($N_{on}$). 
\item Count background events in the signal region ($N_{bg}$). 
\item Determine the number of excess in the signal region ($N_{excess}$\,=\,$N_{on}$\,-\,$N_{bg}$).
\item Determine the effective observation time, effective angular acceptance and finally the energy spectrum.
\end{enumerate}
Selection cuts are applied in order to reject part of the
background.
The \textit{Impact} is selected to be within 10 and 300\,m referred to both telescopes.
The signal region, by means of a cut in the \textit{Hadronness} parameter, is chosen by requiring an acceptance for MC electrons of 60$\%$.
The background distribution is determined through MC proton
simulations. We underline that by now we limited the
production with only proton events. This choice is motivated by the fact
that protons constitute by far the major component of the total CR
spectrum (followed by Helium). On the other hand, we mention that a
faulty description of the background may alter our results
considerably.
This is made even more
complicated by the fact that hadronic interactions are difficult to
simulate with high precision: many types of interactions compete, and not all details
of cross sections are known, particularly at high energies \cite{hadmodel}.
The hadronic showers have been simulated with CORSIKA \cite{corsika},
using the FLUKA model \cite{fluka} for the low energies, in
combination with the QGSJet-II \cite{qgsjet} interaction model for the high energies. MC proton events have
been simulated in a \textit{Zenith} angle range between 5$^o$ and 30$^o$ with
a maximum \textit{Impact} parameter of 1.2\,km in a solid angle of 0.034\,sr.
Since the MC proton simulations have been performed with a different energy spectrum\footnote{to increase the statistics at high energy.}  of $E^{\Gamma_{sim}}$ (with $\Gamma_{sim}$\,=\,-2.0 or -1.78) compared to the real cosmic-ray spectrum of $E^{\Gamma_{real}}$, the \textit{Hadronness} distribution was corrected. Thus, to each event $i$ of the \textit{Hadronness} distribution a weight factor of $w_i$\,=\,$E_{true}^{(\Gamma_{real}-\Gamma_{sim})}$ is assigned; $E_{true}$ is the simulated energy.
In this case, while the uncertainties on the ON data are computed
according to the Poisson statistic,
the uncertainties on the MC protons are defined as $\Delta
N_{pr}$\,=\,$\sqrt{\sum_iw_i^2}$, the square root of the sum of the weights
in the considered bins of the \textit{Hadronness} distribution. The
normalization factor $\alpha$\,=\,$\frac{n_{on}}{n_{pr}}$ is the ratio
between the numbers of ON and MC proton events in the non-signal
region which implies $N_{bg}$\,=\,$\alpha N_{pr}$.
The significance of the excess is defined as:
$$S=\frac{N_{excess}}{\Delta {N_{bg}}}=\frac{N_{on}-\alpha
  N_{pr}}{\alpha\left[\Delta N_{pr}^2+N_{pr}^2
   \left(\frac{1}{n_{on}}+\frac{\Delta n_{pr}^2}{n_{pr}^2}
    \right)\right]^{1/2} } $$
\\ \\ The electrons are simulated in a circular area
of 650\,m radius in a $Zenith$ angle range between 5$^o$ and 30$^o$. The effective acceptance of the telescope system, used for the spectrum
calculations, was considered as the solid angle of acceptance of the
simulated electrons, which in our case corresponds to
$\Omega$\,=\,0.019\,sr. 
\section{Results}
\begin{figure}[tbph]
\begin{center}
\mbox{\includegraphics[width=8cm]{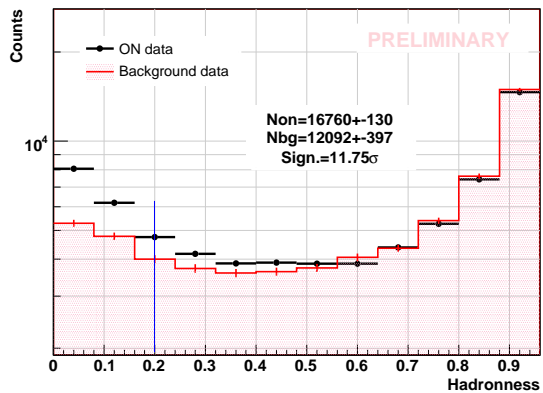}}
\end{center}
\caption{\footnotesize Total \textit{Hadronness} distributions of the ON events (in black) and background simulation (in red) for energy between 150\,GeV and 2\,TeV.}
\label{fig:hadonofftot}
\end{figure}

\begin{figure}[tbph]
\begin{center}
\mbox{\includegraphics[width=6cm]{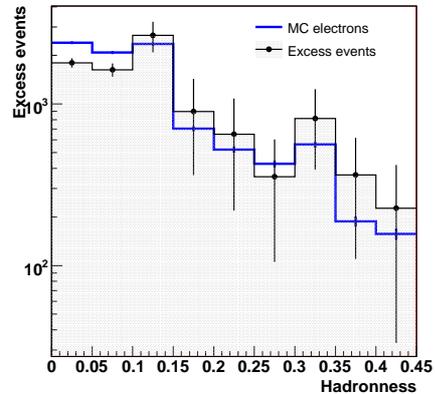}}
\end{center}
\caption{\footnotesize The \textit{Hadronness} distribution of the
excess events (in grey) is compared with the one of the MC electrons for energy between 150\,GeV and 2\,TeV.}
\label{fig:hadexcess}
\end{figure}

\begin{figure}[tbph]
\begin{center}
\mbox{\includegraphics[width=9cm]{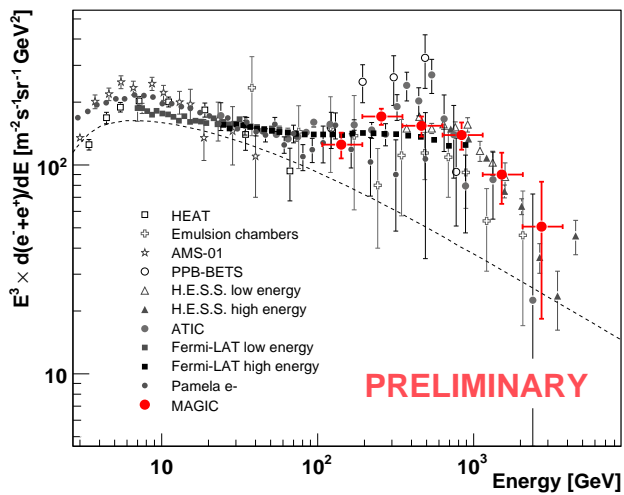}}
\end{center}
\caption{\footnotesize The electron spectrum measured in the energy range between 100\,GeV and 3\,TeV by MAGIC is compared with previous measurements from \cite{heat}, \cite{emulsion}, \cite{ams}, \cite{bets}, \cite{hesslow}, \cite{hesshigh}, \cite{atic}, \cite{fermielectron2}, \cite{pamelael}.}
\label{fig:E3F}
\end{figure}

The total \textit{Hadronness} distribution of the observed events is
plotted together with the normalized distribution of background
events in figure \ref{fig:hadonofftot}. 
The normalization is done in the \textit{Hadronness} range between 0.4 and
0.8. In total, 4668 electron excess events are seen with a
significance of 11.75\,$\sigma$ in the energy range between 150\,GeV
and 2\,TeV. In figure \ref{fig:hadexcess}  the
\textit{Hadronness} distribution of the excess events is instead
compared with that of the MC electron simulations, at low
\textit{Hadronness} values, where the electron signal is expected. The two
distributions, well in agreement, within the systematic uncertainties, validate the obtained result. A statistical test of compatibility in shape between the two distributions is done using the Kolmogorov test. A probability that the excess events follow the distribution of the MC events is found to be more than 70\% in each energy bin.
\\ \\
The spectrum is derived from the measured excess events in the energy range from 100\,GeV and 3\,TeV. A preliminary spectrum can be fitted by a simple power-law with differential index $\Gamma\,=\,-3.16\pm0.06(stat)\pm0.15(sys)$. In figure \ref{fig:E3F} the MAGIC electron spectrum is compared with the measurements from other experiments. It is shown in the form of $E^{3}\frac{dF}{dE}$. The MAGIC spectrum shows some overlap with the direct measurements of ATIC, Fermi PPB-BETS and emulsion chambers and it is in agreement within errors with both the ATIC and Fermi measurements (the bump observed by ATIC cannot be confirmed nor excluded though). At higher energies the MAGIC spectrum overlaps and is in agreement with the measurements of H.E.S.S. .
The spectrum can be improved with higher statistic both in the data and especially in the background simulation.
The measured preliminary electron spectrum suffers from large systematic uncertainties. Main sources of errors are due to atmospheric variations and the atmospheric model used in the simulations, reflectivity and aging of the mirrors, uncertainty in the quantum efficiency, photoelectron collection efficiency and gain of the PMTs. Moreover trigger inefficiencies, camera inhomogeneity and read-out uncertainty provide also systematic effects. Uncertainties due to the hadronic interaction model adopted contribute to most of the errors. Overall, the systematic errors on the flux normalization are estimated to be at the level of 30\,$\%$. The systematic errors on the spectral slope, still under investigation, are more difficult to be estimated and a minimal value expected is $\pm$0.15 \cite{borla}.
\section{Acknowledgements}
We would like to thank the Instituto de Astrof\'{\i}sica de
Canarias for the excellent working conditions at the
Observatorio del Roque de los Muchachos in La Palma.
The support of the German BMBF and MPG, the Italian INFN, 
the Swiss National Fund SNF, and the Spanish MICINN is 
gratefully acknowledged. This work was also supported by 
the Marie Curie program, by the CPAN CSD2007-00042 and MultiDark
CSD2009-00064 projects of the Spanish Consolider-Ingenio 2010
programme, by grant DO02-353 of the Bulgarian NSF, by grant 127740 of 
the Academy of Finland, by the YIP of the Helmholtz Gemeinschaft, 
by the DFG Cluster of Excellence ``Origin and Structure of the 
Universe'', by the DFG Collaborative Research Centers SFB823/C4 and SFB876/C3,
and by the Polish MNiSzW grant 745/N-HESS-MAGIC/2010/0.

\clearpage


\begin{thebibliography}{}
\bibitem{dm} Bi, X.-J., He, X.-G., and Yuan, Q., Phys. Lett., 2009,
  {\bf B678}: 168-173
\bibitem{pulsar} Grasso, D., Profumo, S., Strong, A. W. et al,
  Astroparticle Physics, 2009, {\bf 32}: 140-151
\bibitem{sn} Blasi, P., Phys. Rev. Lett., 2009, {\bf 103}: 051104
\bibitem{hesshigh} Aharonian, F. et. al., Phys. Rev. Letters, 2008, {\bf 101}: 261104-+
\bibitem{fermielectron2} Ackermann, M. et al, Phys. Rev., 2010, {\bf D82}: 092004
\bibitem{atic} Chang, J. et. al., Nature, 2008, {\bf 456}: 362-365
\bibitem{hesslow} Aharonian, F. et. al., aap, 2009, {\bf 508}: 561-564
\bibitem{borla2} Borla Tridon, D., et al, NIM, 2010, {\bf 623}:437-439
\bibitem{rf} {Bock}, R. K. and {Chilingarian}, A. and {Gaug}, M. and {Hakl}, F. and 
	{Hengstebeck}, T. and {Ji{\v r}ina}, M. and {Klaschka}, J. and 
	{Kotr{\v c}}, E. and {Savick{\'y}}, P. and {Towers}, S. and 
	{Vaiciulis}, A. and {Wittek}, W., Nuclear Instruments and
        Methods in Physics Research A, 2004, {\bf 516}: 511-528
\bibitem{hillas} Hillas, M. A., ICRC proceeding, 1985, {\bf 3}:
  445-448
\bibitem{hadmodel} Maier, G. and Knapp, J., Astroparticle
Physics, 2007, {\bf 28}: 72-81
\bibitem{corsika} Heck, D. et al., Forschungszentrum Karlsruhe Report,
  1998, {\bf FZKA}: 6019
\bibitem{fluka} CERN-INFN, http://www.fluka.org
\bibitem{qgsjet} Ostapchenko, S. et al., ICRC proceedings, 2005, {\bf 7}: 135
\bibitem{heat} DuVernois, M. A. et al, apj, 2001, {\bf 559}:296-303 
\bibitem{emulsion} Kobayashi, T., ICRC, 1999, {\bf 3}:61-+
\bibitem{ams} Aguilar, et al, Physrep., 2002, {\bf 366:}331-405
\bibitem{bets} Yoshida, K. O. Advances in Space Research, 2008, {\bf 42}:1670-1675
\bibitem{pamelael} Adriani, O., et al, PhysRevLett., 2011, {\bf 106}:201101-201121
\bibitem{borla} Borla Tridon, D., Ph.D. Thesis, 2011
\end{thebibliography}
\end{document}